# Visual Flow-based Programming Plugin for Brain–Computer Interface in Computer-Aided Design

## Tong Bill Xu[a], Saleh Kalantari*[a]

*[a] Human Centered Design, Cornell University*

Corresponding Author: Saleh Kalantari <sk3268@cornell.edu>

## Abstract

Over the last half century, the main application of Brain–Computer Interfaces (BCIs) has been controlling wheelchairs and neural prostheses or generating text or commands for people with restricted mobility. There has been very limited attention in the field to applications for computer-aided design (CAD), despite the potential of BCIs to provide a new form of environmental interaction. In this paper we introduce the development and application of Neuron, a novel BCI tool that enables designers with little experience in neuroscience or computer programming to: (1) gain access to neurological data, along with established metrics relevant to design; (2) create BCI interaction prototypes, both with digital on-screen objects and physical devices; and (3) evaluate designs based on neurological information and record measurements for further analysis. After discussing the BCI tool's development, the article presents its capabilities through two case studies, along with a brief evaluation of the tool's performance and a discussion of implications, limitations, and future improvement.

# 1. Introduction

Brain-computer interfaces (BCIs) are a set of related technologies that allow computers to analyze neurological signals and relay commands based on those signals to an output device. There is a long history of interest in such technologies, going as far back as the 1970s [1]. A central focus in this research area has been on assisting users who have restricted mobility, often by converting brain activity into computer inputs in ways that mimic the functions of mice or keyboards. For example, [2] used Event-Related Potentials (ERPs) in the medial parietal brain region to control a computer cursor, and [3] did the same using Mu rhythms (8–12 Hz) in the motor cortex. BCI technologies have also been used to control wheelchairs, neuroprostheses, and robotic arms [4–7], and human-robot collaboration [8,9]. Traditionally, most BCI systems rely on clear signals obtained from invasive electrodes inserted into the brain [10,11]. Later studies found that non-invasive electrodes, in particular electroencephalography (EEG), could also be sufficient for the creation of well-functioning BCIs [12]. The field has achieved faster response speeds and better performance in recent years by improving sensors and especially by applying more sophisticated interpretive algorithms to parse the neurological data in real-time.

Advances in sensors and processing technology have also made it possible to create less expensive BCIs based on consumer-level EEG devices, though not without some concerns regarding their overall performance. The most important liability in this area is the accuracy of signal classification, given the reduced numbers of EEG channels and reduced signal quality in commercial devices. Maskeliunas and colleagues [13] found that for recognition of attention and relaxation states, Neurosky Mindwave EEG (single channel) achieved only 22.2% accuracy. Emotiv EPOC (14 channels) performed better at 60.5%, which is still far from ideal. Additional studies have found somewhat better results in applying consumer EEG products to detect specific, prominent features of neurological data, such as the N200 and P300 ERP responses [14] and unidimensional motor imagery [15]. Despite the uncertainty in their broader performance, consumer-level EEG devices have the advantage of greatly extending the potential applications of BCI outside of hospital and laboratory settings, for inexpensive use by the public and/or large-scale data collection in everyday environments.

Many commercial applications of EEG-based BCI are categorized as entertainment [16]. However, there are some borderline cases, such as in the area of neurofeedback, or EEG biofeedback, where metrics calculated from brain activity are used for training, intervention, or rehabilitation [17]. One example is "attention training," in which EEG feedback is used to control the size or movement of an object on a computer screen, to enhance the user's ability to focus. This approach has been used as an inexpensive, home-based intervention for children with attention deficit/hyperactivity disorder (ADHD) [18]. Other common applications include concentrative meditation training and the management of anxiety and post-traumatic stress disorder (PTSD) [19–21]. Neurofeedback-oriented BCIs are interesting from a design standpoint because they do not translate brain activity into a conventional interface such as a screen cursor, but instead use the metrics as a direct form of two-way environmental interaction.

Despite the potential of BCIs for providing a new form of environmental interaction, this technology has received very limited attention in the field of computer-aided design (CAD). There are a few studies that attempted to use EEG signals to provide commands for traditional design software (e.g., a "rotate object" command) by the simulation of traditional keyboard input [22–27]. A few others have used brain signals to estimate imagined shapes [28,29] or to study neural aspects of the design process [30]. A major limitation in this area is that BCIs intended to assist disabled individuals in making computer inputs are still relatively slow, unwieldy, and inaccurate compared to mouse-and-keyboard interfaces, especially for professionals who have extensive training and experience in the use of traditional CAD software.

The current authors believe that a better approach to using brain data in design fields would be as input for parametric and generative procedures. In this cutting-edge workflow, designers work with software algorithms to generate numerous forms within certain constraints (parameters), and then filter, down-select, and refine the resulting design options based on their professional skills and knowledge. Such workflows allow designers to focus on the creative and aesthetic aspects of the product while offloading many of the mundane form-development tasks to the generative software—leading to improved efficiency and a lower cost of design [31,32]. The capabilities of current generative design algorithms have proven useful in optimizing performance measures such as material costs, structural stability [33], and energy efficiency [34]. Using BCIs to provide visualization "seeds" or parameter inputs for generative software may help to further reduce the monotony of computer-aided design. Moreover, since the subsequent generated results will be evaluated and refined by the designers, minor "noise" from the BCI's inaccuracies will be more acceptable compared to previous applications where such inaccuracies could lead to incorrect commands.

Another valuable direction in this area is the analysis of neurological data as a form of user feedback about overall designs or about specific changes in a design variable [30,35,36]. As suggested by Bell and colleagues [37], neural measures such as EEG have the potential to reveal unconscious mechanisms and responses that users are not aware of, and can therefore provide a valuable supplement to traditional self-reported design-testing methods such as surveys, interviews, and focus groups. This approach helps decrease the risk of mono-method bias in design studies (i.e., apparently consistent findings that derive from a shared source of measurement error) [38]. In this context using consumer EEG headsets can also help increase ecological validity and generalizability, as the portable technology allows real-time design response data to be gathered outside of laboratory settings. The measure can also be taken in virtual reality [39–43] and to evaluate important aspects of design such as spatial uncertainty [44]. Supplementing such neurological measurements with BCI functionality can provide tremendous additional value, for example by adjusting design variables in real-time based on users' neurological responses, possibly based on user-created low-fidelity prototypes [40], and thus rapidly homing in on an optimized individual outcome. BCI approaches may also find applications in controlling the settings of finished design products, for example by adjusting environmental temperatures or lighting depending on users' mood, or by changing the ergonomic settings of adjustable furniture to provide better comfort.

In the current paper, we describe the development and application of Neuron, a novel BCI tool that enables designers who have little or no experience with neuroscience or computer programming to: (1) gain access to neurological data, including the band-power measurements of each channel/electrode, along with established signal-pattern metrics that reflect, for example, attention and relaxation states; (2) create BCI interaction prototypes, both with digital on-screen objects and with real-world physical devices; and (3) evaluate designs based on neurological information and record these measurements for further analysis. The goal of this tool is not to replace established design workflows, but rather to expand the creative possibilities of early-stage design ideation, particularly as novel means of input for for parametric and generative design approaches. Neuron can also provide important benefits in terms of granting designers access to neural data to help evaluate human responses to design, in a real-time, interactive format. Thus, Neuron aims to provide enhanced utility and support within the context of existing design workflows. This paper discusses the BCI tool's development, presents its capabilities through two case studies, and provides an evaluation of the tool's performance along with a discussion of implications and future improvements.

## 2. Design

During the design of the tool, we first reviewed the relevant prior technical literature on BCIs and the available software functionality associated with our goals. We then selected an appropriate software platform and created our specific functional modules.

### 2.1. Existing Software Libraries and BCI Tools

Most tools used in the design of BCIs fall into three categories: neuroscience libraries, BCI libraries, and BCI visual programming tools. Neuroscience libraries are created to assist with the analysis of neurological data, usually for research purposes. Examples include EEGLAB [45], FieldTrip [46], and BrainStorm [47], all of which run on the MATLAB platform; as well as the multi-platform library MNE, which has a greater focus on magnetoencephalography rather than EEG [48]. These software libraries require knowledge in the field as well as computer programming skills to use effectively. While neuroscience libraries are not designed specifically for BCI, they are highly useful for providing sophisticated analyses of EEG data as related to various types of mental states or activities. When combined with other functions in MATLAB or Python for EEG feature extraction and communication, all these libraries can be used in BCI applications.

Prominent libraries and software packages created specifically for BCI include BCI2000 [49], BioSig [50], and BCILAB [51]. All of these packages have integrated signal acquisition, processing, and feature extraction functions. Additionally, BCI2000 includes some output functions such as cursor (pointer) and keyboard controls, while BCILAB has implemented a graphical user interface for setting up function parameters. Similar to the neuroscience libraries mentioned above, these packages simplify BCI design, but still require a relatively high level of knowledge in both neuroscience and computer programming.

The final category of relevant tools for our BCI development includes visual programming aids, such as OpenVibe [52]; the EMOTIV plugin for the Node-RED platform [53] [54]; and NeuroPype, which is based on a data-mining suite called Orange [55]. These tools are applications of flow-based programming, which allows users who have limited technical knowledge to tailor software solutions for their specific needs [56–59]. The EMOTIV plugin is contiguous with the EMOTIV company's larger Cortex application, and it grants users access to an impressive array of analytical options, including pre-defined metrics of neurological states and motion-sensor data. Combined with other plugins in the Node-Red system, users with little background knowledge can even set up controls for physical BCI components.

While the capabilities of EMOTIV would be a strong option for our target users, this plugin is currently outdated (it only supports Cortex v.2, not the current version 3 released last year), and the necessary installation of Node-RED can be complex and difficult (it requires the use of the command-line interface for plugin installation). It is also rather complex to integrate Node-RED with standard 3D modeling software used by designers.

Therefore, we decided to use Grasshopper for Rhinoceros 3D as our flow-based programming platform due to its familiarity to designers, its fluid integration with 3D modeling capabilities, and the freedom it provides for ideation, exploration, and form-finding. Another advantage of the Grasshopper platform is its existing user community and library of plugins for data I/O, such as gHowl, and FireFly; machine learning, such as LunchBoxML, and Octopus;

and other purposes. While these plugins were not designed with BCI in mind, they could still support our users to develop the whole BCI process within the platform and provide them with multiple options for similar functions and hence more flexibility and design choices.

## 2.2. Platform Architecture

For the hardware components, we chose OpenBCI, a consumer-level EEG headset, and associated technologies that are open-source and widely regarded as cost-effective [60]. We designed Neuron to focus on data processing, particularly in terms of enabling users to create predictive models with existing machine learning components. We also created a custom interface to read data from OpenBCI devices and send it to additional platforms, which can potentially enable a wide range of applications (Figure 1).

Figure 1. The BCI System Framework

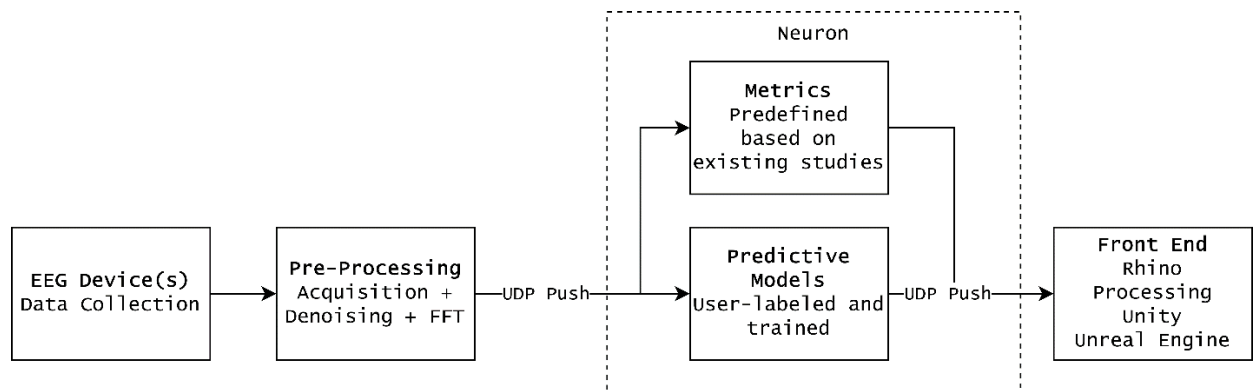

## 2.3. EEG Indicators

The OpenBCI headset (Ultracortex Mark IV) followed the international 10-20 system for electrode placement [61]. The system supports up to 16 channels. The default channel locations are: Fp1, Fp2, C3, C4, P7, P8, O1, O2, F7, F8, F3, F4, T7, T8, P3, P4. The letters indicate the lobe or area of the brain (cortex): Pre-Frontal (Fp), Central (C), Parietal (P), Occipital (O), Frontal (F), and Temporal (T) were covered by the default layout. The numbers further specify the placement with even numbers indicating the right side of the brain, and odd numbers indicating the left side, z (zero, not covered in the default layout) for the midline.

We included six predefined neurological indicators in our current tool (this may be expanded in future work). The selected metrics were based on aspects of neural activities that are of common concern during design processes and in user responses to designs, including Attention, Relaxation, Intention to Move, Workload, Navigation, and Creativity.

Focused *Attention* and its inverse, *Relaxation* (defined as a lack of focused attention), are among the most important aspects of the human cognitive process and have been studied extensively. In prior work, low theta/beta ratio in the frontal area was found to be linked to executive/attentional control [62,63], while alpha power in occipital area was suggested to be linked to the relaxed state, or negatively linked to attention [64,65]. Here we used Beta in Frontal area (F3 and F4) as the approximate indicator of attention, and Alpha in Occipital area (O3 and O4) as the indicator of relaxation. *Intention to Move*, as discussed in the introduction, is linked to the Mu wave (around 8-13 Hz) in the motor cortex [3]. Here we used alpha band (8-12 Hz) at channels in central area (C3 and C4) as an approximation. *Workload* during tasks was suggested to be related to increase in delta and theta power in anterior/frontal area and decrease in beta power in posterior area [66]. Here we used Theta at F3 and F4. *Navigation*, or more specifically, spatial navigation, is possibly linked to theta-band activities in posterior and occipital areas as found in recent studies [67–69]. We selected Theta at O1 and O2 as the indicator. *Creativity*, or creative ideation, was believed to be indicated by alpha synchronization, especially in (pre)frontal, and sometimes posterior parietal area [70]. We included Alpha ta F3 and F4 as the indicator.

It is important to note that the indicators we used in this study are approximations of more precise and rigorous metrics based on lab-level headsets and more advanced data processing. In addition to our pre-defined metrics, the tool has the capacity for users to define their own EEG metrics, as discussed in section 2.4.3 below.

## 2.4. Plugin Components

The BCI software components—including data acquisition (input), signal processing, feature extraction, and visualization output—were created primarily by applying and integrating the functions of existing plugins and libraries. Given our goal of empowering designers to work with the platform to define their own BCI interactions, we used a modular design approach and gave each component only a single function. Figure 2 shows the overall interface for the BCI tool in Grasshopper, and Table 1 summarizes each component and their execution times (these will be discussed in detail in the following sections). In addition, we designed Neuron with respect to other commonly used plugins for Grasshopper that are likely to be used together with our BCI components, such as gHowl, FireFly, LunchBoxML, and Octopus, to avoid unnecessary overlap.

Figure 2. Neuron Components

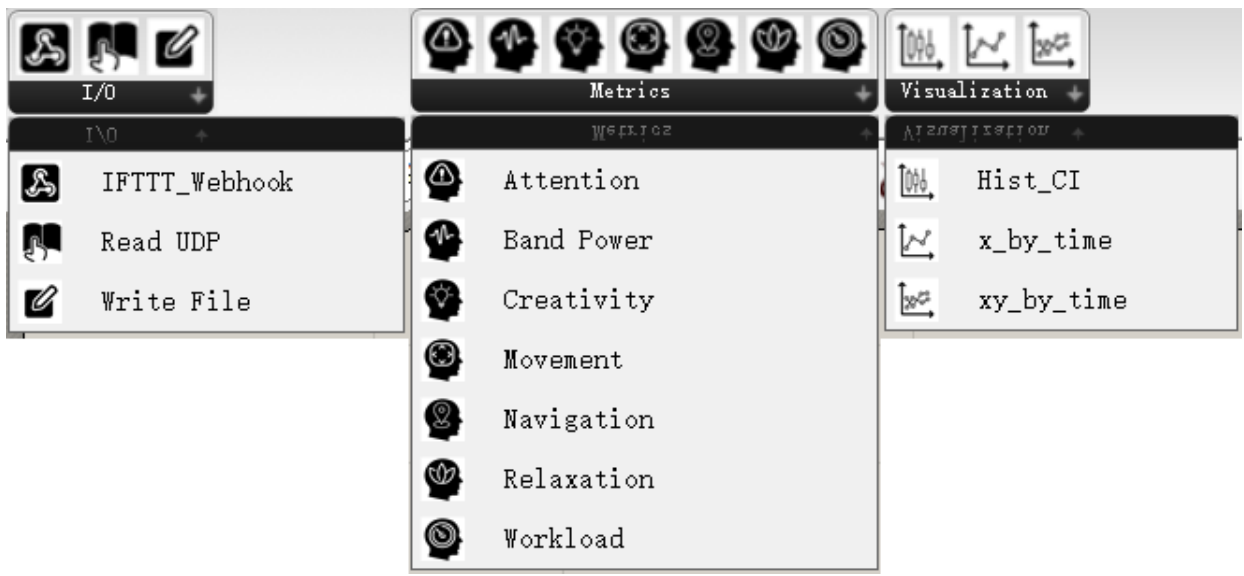

Table 1. Component Descriptions and Execution Times

| Component Name | Description | Execution Time (ms) |
|---|---|---|
| | Data Input/Output | |
| "Read UDP" | Read and parse UDP package, update with timer | 12.5 (2.9) |
| "Write File" | Write incoming data to file with optional timestamp | < 10 |
| "IFTTT_Webhook" | Trigger IFTTT Webhook for Applets | 59.7 (8.4) |
| | Feature Extraction | |
| "Band Power" | Convert FFT data for customized band powers | 35.3 (7.1) |
| other metrics | Calculate pre-defined neural metrics, e.g., Attention | < 10 |
| | Data Visualization | |
| "x_by_time" | Plot data by time | < 10 |
| "xy_by_time" | Plot two data sets by time if both are lists | < 10 |
| "Hist+CI" | Draw histogram and confidence intervals | < 10 |

*Note: Execution time shows the Mean and (Standard Deviation), in milliseconds. Tested with band power data from 16 channels. CIs based on asymptotic distributions.*

### 2.4.1. Data input/output

We used OpenBCI's official interface to handle data acquisition. The software reads data from the headset, performs noise filtering, Fourier transformations, and band-power calculations, and then streams the processed data out through multiple protocols. We selected the UDP protocol for Neuron due to its simplicity and our near-perfect connection quality (achieved since we only needed to communicate with one other program running on the same system). One advantage of this approach is that it greatly reduces execution speed. It is not uncommon for the generative design program and rendering of complicated shapes to take many seconds if not minutes to finish. By leaving time-sensitive raw EEG data outside of Grasshopper (in the OpenBCI interface) and using the processed data instead, we ensured that the intensive form-generation computation would be kept separate from the signal processing. Additionally, leaving the processing to OpenBCI's official software will help ensure that our tool remains compatible with future OpenBCI updates, so long as the data format remains the same.

In Grasshopper, we created a UDP data receiver that can be updated with a timer. This allows users to easily define the desired data refresh rate, and to start or end the auto-update loop with an icon in the system tray. This is very important since setting the refresh interval shorter than the shape-generation and rendering time would essentially freeze the whole platform. After receiving the UDP data, this component unpacks it and converts it into "datatrees", a commonly used structure in Grasshopper that is similar to nested lists.

In addition to our solution, there are two alternative plugins for Grasshopper, "gHowl" and "FireFly," which can read incoming UPD packages. The UDP receiver from FireFly is very similar to our component but has a relatively small pre-allocated buffer for storing incoming data packages; it would be able to read only the band-power data and could not handle the larger Fourier-transform data from OpenBCI. The gHowl UDP receiver, in contrast, continually listens and updates itself with each incoming package. This would usually result in freezing, similar to the timer issue mentioned above (although it could be fixed by adding a "data dam," another native Grasshopper component, which allows for setting predefined update intervals).

Finally, there are multiple options to stream the data from Grasshopper when needed. FireFly includes an Arduino input/output component and a Firmata-like program for running downstream hardware boards, while gHowl has a fully functional UDP sender. We developed a logging component that we simply called "Write File," which outputs the data stream into a CSV file in appending mode (many existing writers for Grasshopper will open the file in writing mode, which erases existing logs and creates a new empty file). CSV (comma-separated values) is one of the most common formats to store tabular data as text files and to exchange data between platforms. We also included an IFTTT component making use of the Webhook interface, which allows users to set triggers for output events with up to three user-definable parameters.

### 2.4.2. Feature extraction for pre-defined metrics

Features of the EEG data can be extracted with formulas or models based on prior empirical knowledge. In the current tool, we have summarized six metrics from previous studies and implemented them as feature extraction components, as described in Section 2.3 above. To evaluate these EEG features, we just need to feed the incoming band-power datatree into the feature extraction component of the tool. One additional step that occurs in this process is rescaling the metrics to a range from 0 to 1, which is done using the Remap Numbers function in Grasshopper. This requires baseline data, gathered by asking the individual wearing the EEG sensors to stay, for example, focused for a few seconds and then unfocused for a few seconds, and recording those periods to identify the 0 and 1 conditions (Figure 3). We also added a band-power calculation function, which allows users to investigate some of the less commonly used EEG bands or waves, such as the Mu wave (roughly 9–11 Hz). This was done by adding the values from corresponding frequency bins together per channel, and then summing band powers from the central lobe/area, C3 and C4 channels on the openBCI headset (Figure 4).

Figure 3. Extracting Pre-defined Features

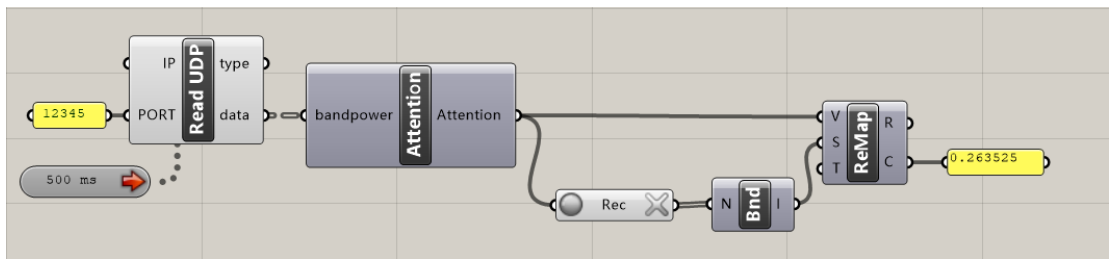

Figure 4. Calculating the Mu Wave on C3 and C4 Channels

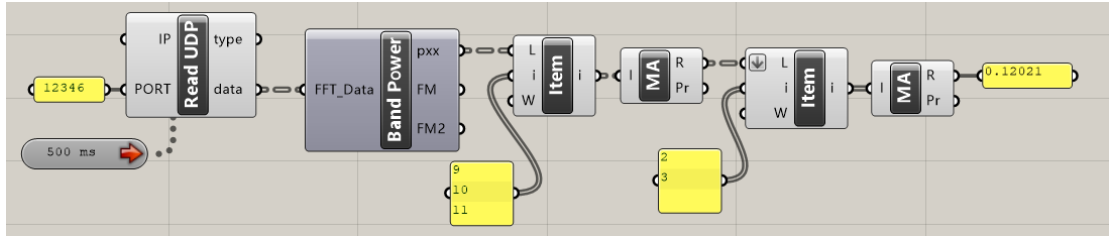

### 2.4.3. Feature extraction for user-defined metrics

Neuron also makes it possible to train new models based on recorded data. Such approaches will likely suffer from sample-size limitations in terms of forming generalizable EEG metrics, but they can be greatly useful for tailoring the BCI to brain activities that vary between individuals, such as motor imagery (in other words, training the BCI to a specific individual's mental imagery). We recommend using the machine-learning components from the "LunchBox" plugin in Grasshopper, which covers most of the algorithms that are useful for classification and prediction in the context of BCIs. Coupled with our "Write File" component, users can easily record brain activities under different conditions to train their own EEG-data classifiers and then use the results to evaluate various types of neural responses to designs.

The first step in this process is to record EEG data under various conditions of neural activity, which could be categorical or continuous. For example, the user might be asked to rate their stress level on a slider from 1–10 (continuous) or to imagine an object moving up vs. moving down (categorical). Figure 5 shows an example of recording EEG band powers into two categorical files for focus states vs. relaxed states. The second step is to use the recorded EEG data to train algorithmic classifiers using machine learning (Figure 6). The classifiers can then be used to sort new incoming data in real-time, thereby predicting whether the users are, for example, in a relaxed state of mind or a focused state of mind (Figure 7). By training models in this fashion for each individual user, it may be possible to achieve better BCI performance. In case a continuous output is desired, the user could similarly train classifiers and use the continuous predicted values instead of classes from the trained models. A better approach would be to record the continuous labels together with the data into the same file and apply algorithms that support continuous labels such as regressions, neural networks, or boosted trees.

Figure 5. Recording Band-power Data into Categorical Files for "Relaxed" vs. "Focused" States

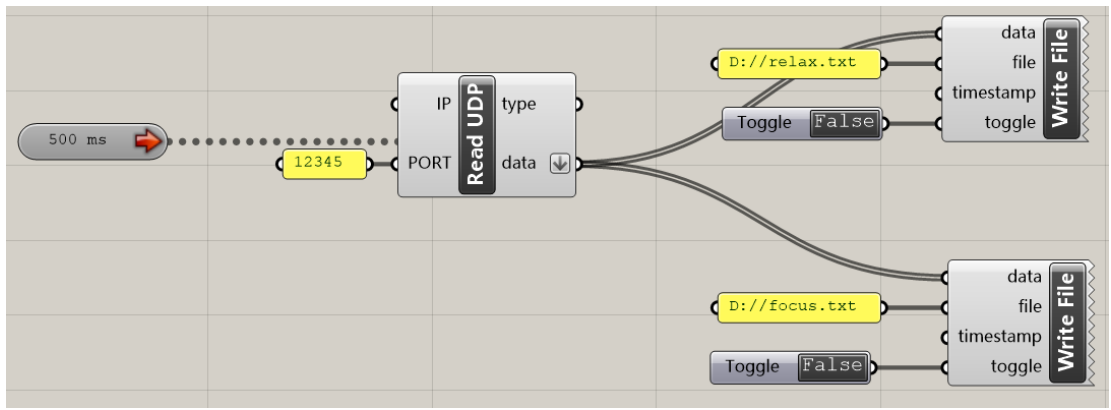

Figure 6. Training a Neural Network Module Using the LunchBox Plugin

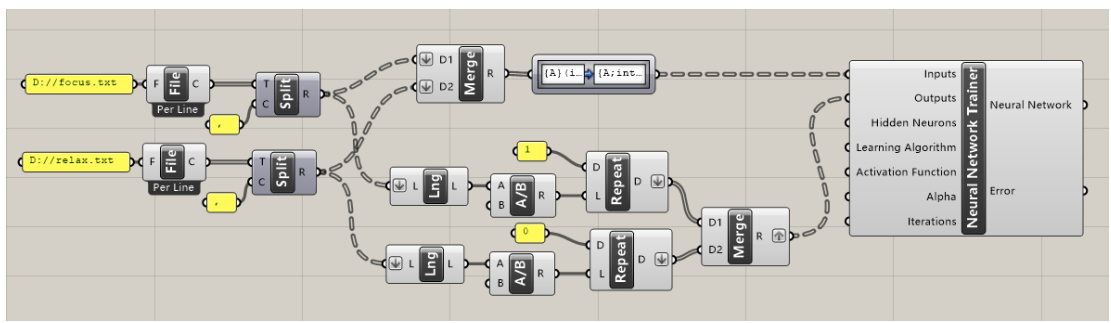

Figure 7. Classifying Incoming Data with a Trained Model (0 Indicates Relaxation States)

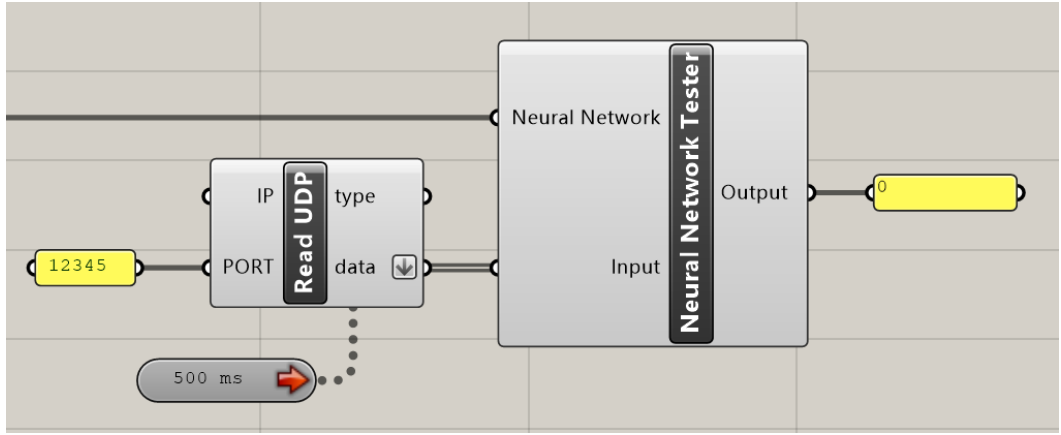

### 3. Case Studies

To evaluate the potentials of Neuron, our new BCI tool, we conducted two case studies in different areas of application: (1) brain-controlled generative design; and (2) a neuro-adaptive smart home device. While the applications discussed here are somewhat rudimentary, they demonstrate the wide range of pathways through which an effective BCI tool can have an impact in design fields, while helping to identify the strengths and weaknesses of our current project. These case studies were not intended to be a scientific evaluation of the BCI's effectiveness, but rather an initial proof-of-concept and exploration of capabilities and usability. Two participants were recruited, one for each case study, using a convenience sampling method. Both participants reported as male, and gave their ages as 29 (Case Study 1) and 32 (Case Study 2). The participants provided oral consent before engaging in the research study, and no personally identifying information was collected. The recruitment procedures and overall study protocols were reviewed and approved by the Institutional Review Board at [removed for the purpose of blind review].

**3.1. Brain-controlled Generative Design (Case Study 1)**

As discussed in Section 1, most existing studies on BCI in design fields have focused on translating brain signals into standard commands for CAD software, or for specific manipulations of shapes. One common feature shared by these studies is the discreteness of the output values, which stems from the types of commands that are used as well as the nature of the classification algorithms applied in the BCI systems. This discrete output leads to undesirable consequences. First, it means that users have only limited ways to interact with the CAD software, as restricted by the interface. Being limited to a relatively small set of pre-defined commands (such as "select circle or square," "rotate the shape," "move shape to the left," etc.) is a significant restriction on creativity that confines users to the set of command possibilities envisioned by the BCI developer. Additionally, the usefulness of such tools for designers who are physically able to employ traditional keyboard and mouse interfaces is highly questionable. In this case (when targeted toward individuals who do not have disabilities), the BCI tool will provide little added functionality, and it may be more cumbersome and less accurate than traditional manual input tools.

For the above reasons, we took a different approach to use BCI in the design process, by focusing on the use of brain data as an "expressive filter" in generative design. The idea is somewhat similar to the "effect pedals" or sliders that are used by musicians to modify the sound of their instruments. By activating one effect pedal or another, the performer can change the overall parameters of the sound in a complex fashion. Although there are other options for expressive controls in generative design software (including simple keyboard switches), we chose this approach because it seemed to be one of the more interesting and effective ways to apply today's "fuzzy" BCI tools (i.e., an area in which precisely mapping conscious user intent to discrete output is not needed and possibly not even desirable). The BCI understood as an expressive filter can directly reflect users' mental states and feelings about a design, thereby promoting intuitive and unhindered expression in the early stages of the form-generation process.

### 3.1.1. Experiment

We trained a new neural metric for a single participant based on mental imagery of "twisting" vs. "straight" objects, following the process discussed in Section 2.4.3. Note that while the machine learning used recorded EEG data only for the two extremes (mentally envisioned twisted states vs. not-twisted), the resulting predicted values were continuous, representing different strengths of twisting depending on how close the user's real-time neural data was to either extreme. We randomly selected 80% of the initial data segments to train the machine-learning model and used the remaining 20% for validation. The ability of the model to predict the data in the validation set was quite strong, showing a statistically significant difference in predicted values between the two conditions (Welch's $t(57.5) = 7.78$; $p < 0.001$) (Figure 8) and a classification accuracy of 84.4% with using a conventional threshold of 0.5 to convert predicted values into two classes. After training the model to evaluate the user's mental imagery, we then connected it to a generative design program as an expressive parameter (Figures 9 and 10).

Figure 8. Validation of the Neural Metric for Visualizing "Twisting" vs. "Straight" Objects

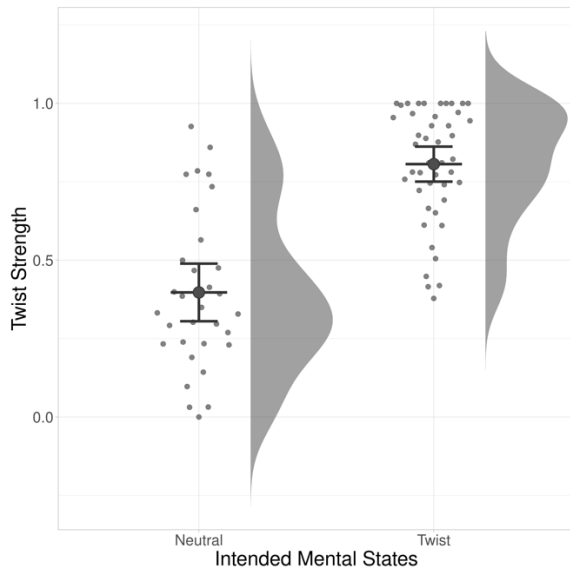

Figure 9. Flowchart of the BCI-influenced Generative Design Process; (at Left) Predictive Model Fitting and (at Right) 3D Model Generation

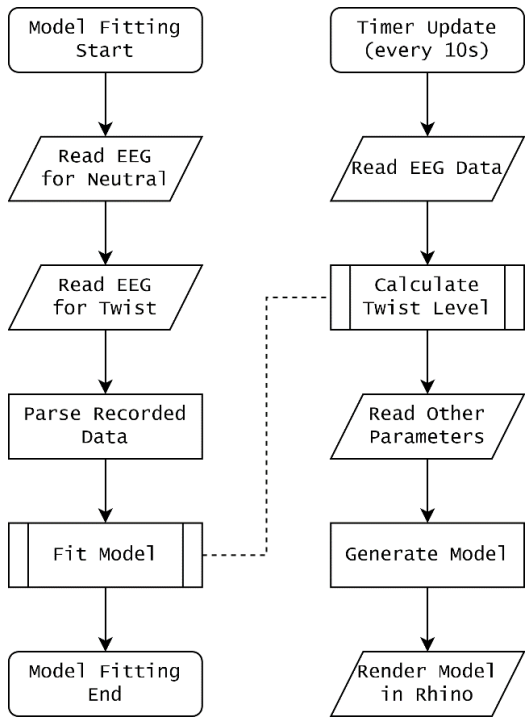

Figure 10. Generative Design Outcomes for Two Structural Geometries (Hexagonal and Triangle) Filtered with the BCI "Twisting" Parameter (Strong "Twist" Mental Visualization Output at Left, and "No-Twist" Output at Right)

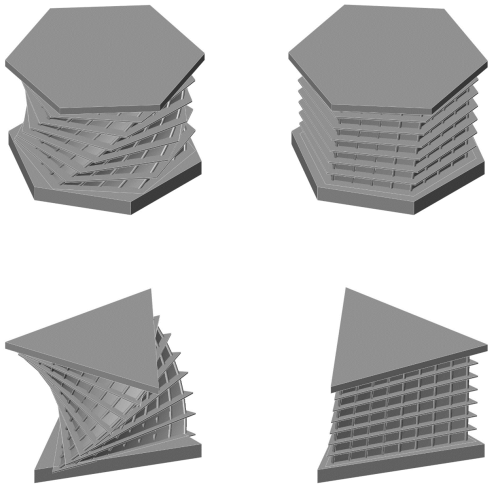

### 3.1.2. Outcomes and Analysis

This case study resulted in a convincing demonstration of the user's ability to apply mental visualizations through the BCI to affect the parameters of generative form creation. Some of the generative parameters of the software were set through standard keyboard commands, most notably the fundamental geometry of the building footprint (triangle, square, hexagonal, etc.). The degree of "twist" in the architecture was treated as a continuous parameter shaped by the BCI output. The user found the system to be effective and reasonably responsive to the mentally visualized desire for "more twist" or "less twist" in the computer-generated designs.

During the experiment, we found several notable limitations and insights. First, while conducting the statistical validation analysis we realized that temporal autocorrelation in the training data [71] may lead to an overestimation of model performance. This issue can be resolved by using data from a separate session for validation, rather than using a random selection of validation data taken from the same session as the model training. Second, motion artifacts in the EEG data were more common than we expected, resulting from the participant's eye and upper-body movements during the experiment. This is a common issue in BCI studies [72], and it can affect the performance of the predictive model. While we did implement standard filtering and smoothing functions in the OpenBCI software intended to remove motion artifacts, these algorithms are limited, and they need to be improved to create better BCI performance in real-world scenarios. Finally, the generation algorithm that we used in the experiment turned out to be relatively slow, taking around 5–9 seconds to finish producing each new revision of the geometry. We had to use an accordingly slow refresh frequency to read the EEG data and calculate the twisting strength to avoid freezing the Grasshopper platform. This latency is too high for an optimal user experience, in terms of fluidly translating mental desires for design changes into the 3D output. To solve this problem, it may be necessary to use coarser (and thus faster) representations of the 3D architecture. Adjusting the tool so that it only execute the generation algorithm at the user's request (rather than together with the EEG data) may also help to improve the fluidity of the BCI experience.

### 3.2. Neuro-adaptive Environment (Case Study 2)

The use of interactive or "smart" technologies within living and working spaces is becoming increasingly common. Such devices can have benefits not only in improving building performance and sustainability, but also in providing an enhanced user experience and a more fluid real-time interface between humans and their built environments. BCI is one of the new technologies that have been explored as a means towards enhancing environmental control [73]. With this approach, designers aim to enable humans to more effectively manage environmental factors such as thermal conditions, lighting, sound, views, or furniture layout, with the ultimate goal of augmenting human performance and everyday functioning. Considering this growing application area for BCIs, we defined our second case study to focus on the use of Neuron for controlling an interactive window curtain that opens or closes based on the user's neurological state. This application of Neuron is quite different from its potential use in providing inputs to CAD software, but it demonstrates how the same tool can be leveraged in a use-case or design-evaluation scenario, with the potential to provide valuable data about human needs and human–environment interactions.

There is a great deal of research literature indicating that exposure to natural light and natural views has a restorative and relaxing effect, improving mood and the subsequent ability to engage in mental effort [74,75]. However, such views can become distractions when it comes time to direct focused attention on a task, leading to poorer performance and lower satisfaction [76,77]. The ability to adjust one's physical environment can therefore be important in maintaining an optimal balance between focused vs. restful attention states over time [78,79]. One solution to help address this need is an adaptive environment that responds automatically to users' mental states.

### 3.2.1. Experiment

We used Nueron's included neural metrics of Attention (focus) and Workload to control a window curtain in an office setting. Trigger events for opening or closing the curtain were set using the IFTTT Webhook interface. An instruction to "open" was sent to a commercial motorized curtain [80] when the EEG Attention metric dropped below a certain threshold, inviting the user to take a short break before going back to work. The trigger to "close" the curtain was attached to the Workload metric so that it would close when the Workload was above a designated threshold (Figures 11–13). The particular threshold levels for these triggers could be adjusted by the user to optimize the results.

Figure 11. Flowchart for the BCI-based Neuro-adaptive Curtain; EEG Threshold Settings for "Low Focus" and "High Workload" Can be Adjusted by the User

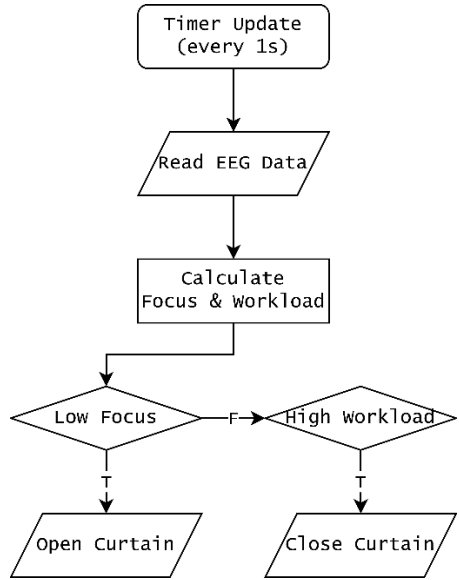

Figure 12. The Responsive Curtain Creates a Human-in-the-loop Adaptive Environment

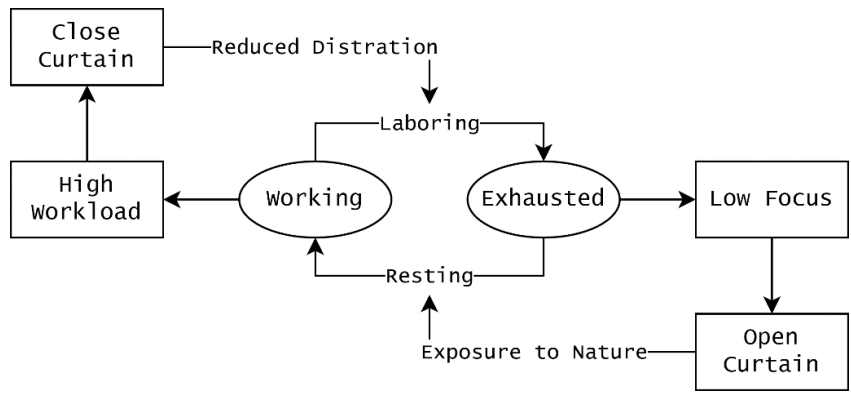

Figure 13. The Neuro-adaptive Curtain Opens When Attention is Low (at Left), and Closes When Workload Is High (at Right)

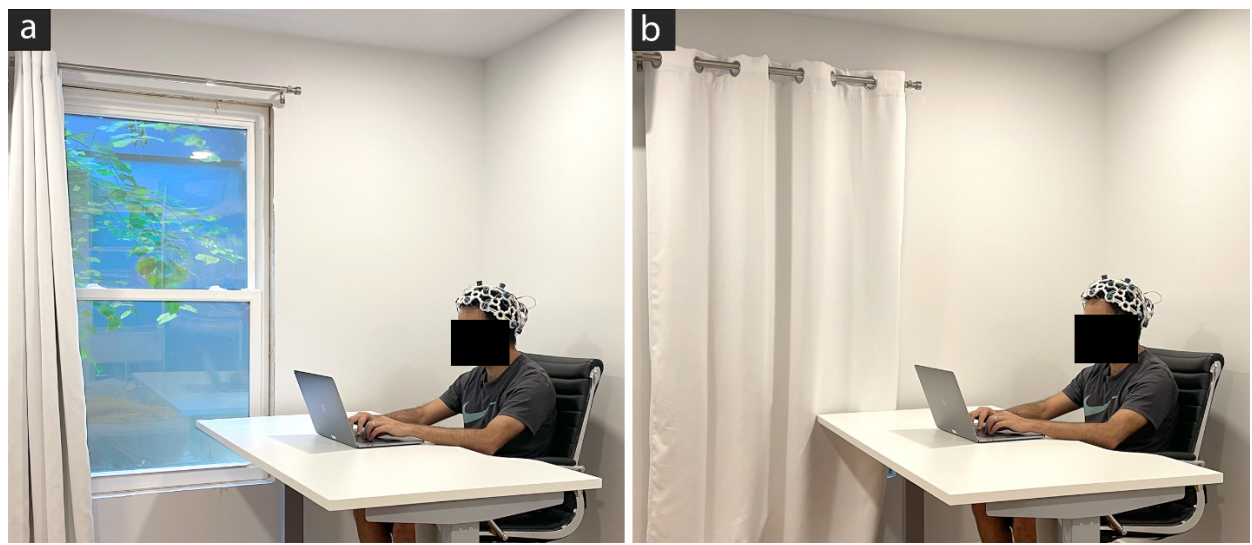

### 3.2.2. Outcomes and Analysis

The BCI was effective in responding to the user's mental states, and its outcomes had a noticeable impact on the room's atmospherics. The underlying concept in this experiment was drawn from the broader design area of intelligent control systems in the built environment, which transform in response to external environmental conditions and/or the needs of users [81–83]. In this case, the adaptive design focused on human neurological functioning, which is a somewhat novel performance condition in adaptive building design. Such approaches are most likely to find a place amid other monitoring devices intended to promote comfort, such as those that automatically respond to changing light levels, thermal conditions, air quality, and so forth [84]. Eventually, an integrated system may be able to learn individual users' behavioral patterns and thereby estimate their neural conditions based on other types of sensors (extent of motion or fidgeting, gaze direction, etc.), so that the continual use of EEG would no longer be required.

## 4. Discussion and Conclusion

The case studies described here show the wide array of potential applications of Neuron, which can be used by designers to efficiently read brain data, extract data features (either pre-defined or customized), and develop specific BCI interface applications. We found that the system met our goals and that all components were stable with acceptable execution times and performance. While the time required to process complicated shape-generation algorithms created a latency effect in some applications, these issues are not inherent to the current BCI software, and they can potentially be resolved by reducing the level of detail in the 3D models.

One of our goals in creating this tool was that users with expertise in various areas could potentially expand its functionality to meet their needs. For example, those who are knowledgeable about EEG systems will be able to define their own neural metrics based on prior literature or recorded data, while non-specialists can just use our included standard metrics. Similarly, users who are experienced with interactive BCI devices can leverage our platform as part of developing more complicated and interesting interactions with devices by exporting the CSV output data. Likewise, researchers with statistical backgrounds may be interested in exporting the CSV output to analytical software packages for conducting robust hypothesis testing. Meanwhile, those without specialized knowledge and research interests can still use the tool for preliminary studies to gain design feedback, and/or to develop simple BCI prototypes and explore their use in the idea-generation process.

Compared to existing BCI tools [45,48,51,52,49], Neuron does not require users to fully understand neuroscience or coding. However, it does require at least a minimal familiarity with Grasshopper's flow-based visual programing language. This software is relatively easy to learn, and it is extremely common in design fields, which is one of the main reasons we chose the Grasshopper platform. Particularly in the area of generative design, Grasshopper is immensely popular, with an established ecosystem, multiple active communities, and various courses and tutorials both online and at universities. Some researchers have argued that flow-based programming, due to its visual nature, might be easier for many designers to understand and implement compared to traditional coding [59]. Thus, we can safely assume that most of our target users will already know, or at least be able to learn, how to effectively use this platform.

Most EEG studies in the industry focused on worker's experience, for example, the workload and attention of construction workers [9,85–87], to prevent hazard [88,89]. The foci were slightly different in the design field, where studies cared more about the design process, especially creativity, and more importantly in the context of our study, evaluation and aesthetic appreciation [90]. Researchers have found that gaps in terminology and language comprehension, particularly in the area of architectural design, may limit the ability of designers and users to communicate their concerns to each other in an effective fashion [91]. Objective data from EEG metrics can assist with this issue, as it has an unparalleled ability to reveal the internal states of users that might otherwise be obscured in self-reported responses or even in behavioral observations. By triangulating EEG metrics against more traditional forms of user testing and feedback, design researchers can reach stronger conclusions and conduct more fine-grained analyses.

An important feature of Neuron is that it can provide support for both the generation of ideas in the early design stages and the later evaluation of completed designs. Access to brain data and associated metrics is an incredibly valuable and flexible tool for multiple aspects of the design process. In the context of idea generation, it is notable that unlike most previous BCI studies [23,24,26,27], we envision these neuro-informed functions to be combined with conventional keyboard and mouse interfaces, rather than replacing them completely. This provides a wide range of options for how designers may choose to integrate BCI functionality into their creative workflow. In terms of design research and evaluation, adding neurological measures and neuro-feedback, in coordination with traditional interviews, surveys, and behavioral observations, can assist in reaching more robust and well-supported conclusions. Since it is possible for the BCI tool to continuously read neural metrics from users, it also has the potential to optimize real-time feedback processes such as evolutionary algorithms or generative adversarial networks, which seamlessly integrate design creativity-and-response cycles.

## 4.1. Expressiveness and Playfulness in Design

One major concern for BCI applications that are not focused specifically on people with disabilities is that they may fail to reach the precision of other input devices. Perhaps, however, there is no need for such a comparison. The types of BCI design generation envisioned in the current project leave room for flaw tolerance, creative randomness, and subsequent revision. After all, ambiguity and the effects of randomness are not necessarily counterproductive in the world of the arts, nor in the scope of natural evolution, and they may also have a role in the practice of design. To return to the example of musical effect pedals, the transformative effects of these filters are often hard to predict for specific melodies. The art form encompasses a large array of such ambiguities, ranging from current directions in expressive MIDI electronics for digital music to the subtle variations expressed by traditional wind instruments and fretless string instruments [92,93]. Similarly, the design need not always function as a precise and measured expression of intended outputs, especially in the initial idea-generation phases of a project.

The interpretation of architecture as a localized, temporal experience and as a continuing process of transformation has also started to gain attention in recent years. Terms such as "weak architecture" have been coined to discuss this sense of transience and mutability [94]. It is also common today to understand design as a form of communication, which is to say, symbols or gestures (in space) that convey an outlook or experience [95]. In this sense, expressive BCIs can help enable designers to engage with the creative process by more directly linking their own neurological reactions and mental states to the evolution of the design product. With such considerations, the use of BCI in design should be understood as going beyond mere "entertainment" [16], but we may also recognize it as willfully encompassing the role of "play" in the development of our aesthetic visions and civilizational goals [96].

## 4.2. Limitations and Future Research

The accuracy of BCIs will continue to be an issue of concern, even when such technologies are oriented toward "playful" and creative output. In our Case Study #1 the accuracy for introducing "twist" into an architectural model was quite respectable at 84.4% for the two-category comparison, falling within the higher range of previous BCI studies [97–99]. Some researchers have reported achieving nearly 100% accuracies, though it is likely such findings are due to lapses in study designs, such as those leading to temporal autocorrelation or entailing very small validation sample sizes [71]. The ongoing development of more sensitive consumer EEG sensors, as well as more robust de-noising algorithms and machine-learning capabilities, should help in continuing to improve the effectiveness of these technologies.

In the broader research context, the use of predictive models with no underlying causal hypotheses has also been criticized [100]. Machine-learning models can have a striking and sometimes almost uncanny utility, but there is a legitimate concern about the extent to which these technologies, which are based on pure number-crunching to find correspondences, actually expand human knowledge and lay the groundwork for thoughtful action. While we followed similar approaches in our own brief case studies, we agree that much more work needs to be done to place BCI research onto a stronger empirical footing. It is also worth noting that the prediction accuracy of this and many similar studies may not be representative of real-world scenarios, due to the wide array of confounding variables that may be present in organic environments and that may have an impact on neurological responses. The golden standard for testing the usability of BCI tools is measuring the time needed for users to reach their goals during everyday use, as can be seen in various BCI movement-control studies [12]. We hope to reach this point of everyday-use testing for Neuron in the future, but the current product is still in the laboratory evaluation and revision stage.

For the future technological development of Neuron, one of the goals is to make it compatible with a wider array of hardware devices, extending beyond the OpenBCI family. We have recently added the ability to retrieve TCP data packages from the OpenVibe BCI system and thus read data from all of their supported devices. However, a dedicated interface with its own data pre-processing functions would make the system more fluent and more translatable to other EEG hardware inputs. We also intend to expand the range of pre-defined EEG metrics beyond the six that are currently available (measures of Attention, Workload, etc.). While knowledgeable users of Neuron have free reign to define their own metrics as frequency ranges, and to build their own predictive models for specific neurological features, the addition of more pre-defined metrics based on prior EEG studies will make it easier for those with little background knowledge in neuroscience to use the tool to investigate specific design outcomes.